\documentclass[fleqn,usenatbib]{mnras}

\usepackage{graphicx}
\usepackage{txfonts}
\usepackage{xcolor}
\usepackage{dblfloatfix}
\usepackage{amssymb}	
\usepackage{multicol}  
\usepackage{pdflscape}	

\usepackage[T1]{fontenc}

\DeclareRobustCommand{\VAN}[3]{#2}
\let\VANthebibliography\thebibliography
\def\thebibliography{\DeclareRobustCommand{\VAN}[3]{##3}\VANthebibliography}

\usepackage{graphicx}	

\newcommand{\ARAA}{Annual Review of Astronomy and Astrophysics}
\renewcommand{\textbf}[1]{#1}

\title[Radio emission impact on SMBH accretion rates]{The effect of extended radio emission on SMBH accretion rate estimates}

\author[S. Amarantidis et al.]{
		   Stergios Amarantidis,$^{1}$\thanks{E-mail: samarant@iram.es}
			Duncan Farrah,$^{2}$
			Nick Seymour,$^{3}$	
			Mark Lacy,$^{4}$		
		    Iris Breda,$^{5}$
            Bodo Ziegler,$^{5}$
			\newauthor \, Olmo Piana,$^{6,7,8}$
			Miguel Sánchez-Portal$^{1}$        
\\
$^{1}$Institut de Radioastronomie Millimétrique (IRAM), Avenida Divina
Pastora 7, Local 20, E-18012, Granada, Spain\\
$^{2}$Department of Physics and Astronomy, University of Hawai’i at Mānoa, 2505 Correa Rd., Honolulu, HI 96822, USA\\
$^{3}$International Centre for Radio Astronomy Research, Curtin University, Turner Avenue, Bentley, WA 6102, Australia\\
$^{4}$National Radio Astronomy Observatory, 520 Edgemont Road, Charlottesville, VA 22903\\
$^{5}$University of Vienna, Department of Astrophysics, Türkenschanzstraße 17, 1180 Vienna, Austria\\
$^{6}$Department of Physics, Informatics and Mathematics, University of Modena and Reggio Emilia, 41125 Modena, Italy\\
$^{7}$Department of Physics, National Taiwan Normal University, No. 88, Section 4, Tingzhou Road, Taipei 116, Taiwan, R. O. C.\\
$^{8}$Centre of Astronomy and Gravitation, National Taiwan Normal University, No. 88, Section 4, Tingzhou Road, Taipei 116, Taiwan, R. O. C.
}

\date{Accepted 2025 October 06. Received 2025 September 13; in original form 2025 April 30}

\pubyear{2025}

\begin{document}
\label{firstpage}
\pagerange{\pageref{firstpage}--\pageref{lastpage}}
\maketitle

\begin{abstract}
Accretion rates in radio galaxies are typically estimated from optical and total radio flux measurements, incorporating emission from the core, jets, and lobes. These estimates can be used to investigate the link between observed Active Galactic Nuclei (AGN) emission properties and the underlying accretion physics of their Super-Massive Black Holes (SMBHs). However, while optical and core radio emission trace the ongoing accretion episode, extended jet and lobe structures may result from past AGN activity. Therefore, accretion rates inferred from spatially unresolved radio observations may be systematically overestimated, a bias whose prevalence and extent have yet to be thoroughly explored. In this study, using a sample of 121 local radio-loud galaxies with spatially resolved radio components, we assess this effect by estimating their \textit{Eddington}-scaled accretion rates ($\lambda$) using both the common methodology which considers total radio fluxes and a simple but novel approach that treats core and extended emission as signatures of distinct accretion phases. Our results show that the former method systematically overestimates the $\lambda$ by a factor of $\sim 3$, affecting the accretion mode classification in approximately $11\%$ of sources. This discrepancy appears to correlate with radio size, with the most extended galaxies indicating a transition in accretion disk mode. Such a bias could affect AGN classification in unresolved high-redshift radio surveys. Our results motivate re-examining accretion rate calculations from AGN radio surveys and align with the AGN unification model for radio galaxies, revealing a clearer link between accretion disk physics and optical spectral properties.
\end{abstract}




\section{Introduction}
A long-standing challenge in the field of extra-galactic astronomy revolves around the complexities associated with the emission produced by Active Galactic Nuclei (AGN) and the underlying physics governing Super-Massive Black Holes (SMBHs). Due to the intricate nature of the composition and dynamics within accreting SMBHs, observations of different regions within the accretion disk can give rise to AGN with diverse and distinctive physical characteristics. In this context, AGN can be categorised based on observables, such as their detection wavelength, emission lines, or luminosity \citep[for an extensive review see][]{2017A&ARv..25....2P}. For instance, at optical wavelengths AGN are categorised into two primary types, Type 1 and Type 2, depending on the observer's line of sight relative to the accretion region and the presence or absence of broad emission lines in their spectra \citep[e.g.,][]{2014NatPh..10..417V}. Furthermore, AGN are classified as either High-Excitation Radio Galaxies (HERGs) or Low-Excitation Radio Galaxies (LERGs), a classification based on the intensity of their optical emission lines. Specifically, HERGs are characterised by displaying flux ratios of $\rm [OIII]/H_{\alpha} > 0.2$ and equivalent widths of $\rm [OIII] > 3 \, \AA$, whereas LERGs exhibit relatively weaker [OIII] fluxes \citep[e.g.,][]{1994ASPC...54..201L}. 

Additionally, at radio frequencies, AGN can be categorised according to the ratio of their radio-to-optical flux as radio-quiet/loud \citep[e.g.,][]{2021PASJ...73..313Z}, or according to the morphology of their mechanical output. In this sense, galaxies with extended radio jets (i.e., length > 15-20 kpc) can be classified as \textit{Fanaroff-Riley} I or II \citep[FR; see][]{1974MNRAS.167P..31F} based on the location of their peak radio luminosity: FRI sources exhibit core-dominated emission near the host galaxy, whereas FRII sources display lobe-dominated emission at greater distances from the core.

In the X-ray regime, AGN typically display a `universal' emission pattern characterised by high luminosity (i.e., $L_{\rm 2-10keV}>10^{42} \, \rm erg/s$) across the entire X-ray energy spectrum. This X-ray emission is believed to emanate from the inner disk of rapidly accreting SMBHs in what is called quasar or radiative mode \citep[for an extensive review see][]{fabian12}. Typical values of this mode are above 1$\%$ of the \textit{Eddington} accretion rate, a limit that corresponds to an accretion state where the radiation pressure generated by the accretion process equals the gravitational force at all radii \citep[see][]{2007MNRAS.374.1146H}. Such fast accretion produces a geometrically thin, optically thick disk (TD; \citealt{Shakura1973}) which leads to strong X-ray emission due to inverse-Compton interactions with electrons in the hot corona around the SMBH. A second accretion mode, often referred to as the radio or jet mode, operates typically at low accretion rates (below $1\%$ of the \textit{Eddington} limit) and is characterised by an Advection-Dominated Accretion Flow (ADAF; \citealt{rees1982}). In this mode, the flow is commonly associated with two bipolar outflows of material, known as jets, where the potential energy of the infalling matter is transformed into kinetic energy.

Notably, estimating the accretion rate and classifying a radio galaxy as either quasar-mode or radio-mode remains a challenging task, typically relying on the integration of the total radio flux, which includes contribution from the core, jets, and lobes of the galaxy. In this study, we propose a novel methodology for the estimation of the accretion rate of radio galaxies, motivated by the non-trivial realisation that the optical, UV and X-ray emission of an AGN correspond to the ongoing accretion disk and its associated accretion rate, while the extended radio jets might originate from a past active SMBH accretion event. 

This distinction becomes particularly evident in the case of powerful radio galaxies where large-scale radio jets are frequently observed extending over tens of kiloparsecs \citep[e.g.,][]{2022JApA...43...97S}, when resolution is limited \citep[e.g., in the case of MHz observations;][]{2022MNRAS.513.3742K}, or when dealing with high-redshift sources with small angular sizes. The formation of such extended jets, even under the assumption of relativistic velocities for the ejected material, typically requires several million years of continuous expansion. Given that variations in accretion rates have been both observed and modelled on much shorter timescales \citep[e.g.,][]{2011ApJ...737...26N,2021ApJ...909...82R} it becomes apparent that the `older' accretion disk, responsible for the radio jets, may not share the same physical properties (e.g., accretion rate, geometrical thickness) as the `current' disk responsible for the X-ray, optical and core radio emission.

Over the years, the exploration of various AGN manifestations and SMBH properties has led to the noteworthy observation that many of these objects exhibit similarities in their observables, indicating that they could represent the same entity with different structural characteristics \citep[e.g.,][]{2017A&ARv..25....2P,2021IAUS..356...29S}. For instance, a blazar has been revealed to be a radio-loud AGN, similarly to the FRI and FRII radio galaxies, with its radio jet oriented face-on to the observer. In the same manner, a broad-line AGN is distinct from a narrow-line radio galaxy depending on the corresponding gas-cloud region of the accretion that is observed. Additionally, optically identified AGN as Type 1 are commonly unabsorbed at the X-ray energies, while Type 2 AGN seem to have higher obscuration factors (i.e., $N_{\rm H} > 10^{22} \, \rm cm^2$). 

Such a unified scenario presents an especially intriguing prospect when examining radio galaxies, with their sub-categorisation raising the compelling question of whether these objects reflect fundamental distinctions in the physical properties of SMBHs and their corresponding accretion disks. In this regard, it has been shown \citep[e.g.,][]{2012MNRAS.421.1569B,2021ApJ...911...17G} that most HERGs accrete at the quasar mode regime, whilst the LERGs are associated mainly with the radio mode \citep[e.g.,][]{heckman14,2022A&A...662A..28M}. Additionally, it seems that most FRI AGN are also LERGs, while FRIIs tend to exhibit strong optical emission lines similar to the HERGs \citep[see][]{2020NewAR..8801539H,2021A&ARv..29....3O}. Nevertheless, such relation is not straightforward, with recent studies demonstrating that low-luminosity FRII galaxies could also be classified as LERGs \citep[e.g.,][]{2022MNRAS.511.3250M} or that HERGs could accrete at the radio mode regime \citep[e.g.,][]{2022MNRAS.516..245W}. In this context, it becomes crucial to re-evaluate the AGN unification model by applying the aforementioned methodology to a sample of radio galaxies, assessing the role of past versus present accretion events in shaping the observed connection between H/LERGs and their accretion modes.

This article is structured as follows: in section 2 a modified methodology of estimating the accretion rate parameter for SMBHs is presented, while section 3 is dedicated to the presentation of the results with a brief discussion of their impact. Lastly, the conclusions and potential future advancements in the field are discussed in section 4. Throughout this paper, a $\rm \Lambda$CDM cosmology is assumed with $H_0 = 70\, \rm km s^{-1} Mpc^{-1}$, $\Omega_{\rm m} = 0.3$, and $\Omega_{\Lambda} = 0.7$.

\section{Methodology}
To address this issue and quantify its significance, we initially apply our methodology to a well-studied, resolved, local radio galaxy and subsequently to a sample of FR galaxies.

\subsection{Example extended radio galaxy} 
A common approach to estimating the accretion rate of a radio galaxy, as compared to its \textit{Eddington} luminosity ($L_{\rm Edd}$), is to adopt the methodology presented in studies such as \citet{2012MNRAS.421.1569B}. In this approach, the accretion rate is derived by the sum of the radiative/bolometric luminosity, which is based on the optical $\rm [OIII]$ emission line (i.e., $L_{\rm bol,curr}=3500 L_{\rm [OIII]}$; \textnormal{obscuration and stellar emission might contaminate this luminosity}; see \citealt{heckman14}) and the mechanical radio output. The latter is determined by the empirical relation \citep[][]{2010ApJ...720.1066C}: 
\begin{equation}
L_{\rm mech,total}=7.3 \times 10^{36}\left(\frac{L_{\rm 1.4GHz,total}}{10^{24} \, \rm W/Hz}\right)^{0.7}\, \rm W, 
\label{equation-1}
\end{equation}
where $L_{\rm 1.4GHz,total}$ is the 1.4 GHz luminosity of the source including its total emission (i.e., core, jet and lobes). In this regard, the \textit{Eddington}-scaled accretion rate, as it would have been presented in the literature, is estimated as follows:
\begin{equation}
    \lambda_{\rm liter} = \frac{L_{\rm total}}{L_{\rm Edd}} = \frac{L_{\rm bol,curr}+L_{\rm mech,total}}{L_{\rm Edd}}.
    \label{equation0}
\end{equation}

Hercules A (or 3C 348; Ra, Dec: 252.783286, +04.993208) is one of the FRI, radio-loud, low-z galaxies (i.e., $z=0.1549$), observed across multiple wavelengths, that presents extended, bright radio jets and lobes of about 153 kpc \citep[e.g.,][]{2003MNRAS.342..399G}. With a SMBH mass of $M_{\rm SMBH} \sim 2.5 \times 10^9 \, \rm M_{\odot}$ \citep[estimated from the K-band luminosity of the bulge; see][]{2006ApJ...652..216R,2016PASJ...68...26F}, the \textit{Eddington} luminosity of this source is equal to: $L_{\rm Edd} = 4\pi G M m_{\rm p} c/\sigma_{\rm T} = 1.26 \times 10^{38} \left(\frac{M_{\bullet}}{M_{\odot}}\right) = 3.15 \cdot 10^{47} \, \rm erg/s$. According to \citet{2022A&A...658A...5T} its total radio flux at 1.4 GHz is equal to 51.5 Jy, while it presents an optical luminosity of $L_{\rm [OIII]}\sim 10^{40.9} \rm \, erg/s$ \citep[see][]{2021A&A...645A..12B}, leading to a $\lambda_{\rm liter}$ value of 0.07.

In contrast, the methodology proposed in this study considers two distinct accretion episodes: a past event responsible for the extended radio jets and lobes, and a current accretion disk driving the observed radio and optical core emission. In this regard, having in mind the resolved nature of the radio observations of Hercules A, we can first consider solely its central radio emission ($L_{\rm mech,core}$) and subsequently its combined jet and lobe mechanical power ($L_{\rm mech,past}=L_{\rm mech,jet}+L_{\rm mech,lobes}$). The latter value can be estimated by the aforementioned empirical relation (equation \ref{equation-1}) connecting the mechanical output of a radio jet with its total radio emission and is typically derived from observations of X-ray cavities surrounding the radio lobes. 

On the other hand, the kinetic output of the currently observed radio core of the galaxy (i.e., unresolved radio jets) on the intergalactic medium is unknown. Nevertheless, we approximate this impact using the monochromatic radio core emission by \citet{2007ApJ...658L...9H}:
\begin{equation}
L_{\rm mech, core} = W_{0}\left(\frac{L_{\rm radio, core}}{L_{0}}\right)^{\frac{1}{1.42-a_r/3}},
\label{equation3}
\end{equation}
where $W_0$ is a normalisation parameter, $L_{\rm 0}=7 \times 10^{29} \, \rm erg/Hz/s$ and $a_{\rm r}$ is the radio spectral index. Considering the unknown nature of $W_0$, we can approach the estimation of $L_{\rm mech,core}$ with a range of values indicated by \citet{2007ApJ...658L...9H} adopting $W_0=4 \times 10^{43}-1.3 \times 10^{44}\, \rm erg/s$. While this range is derived by X-ray cavity observations, possibly corresponding to a past accretion event, we adopt it in our analysis to estimate the mechanical feedback from the core jet (a more detailed discussion of this topic can be found in sub-section \ref{caveats}). Therefore, with a flux of $f_{\rm 1.4 GHz, core} \sim 39.2 \, \rm mJy$ \citep[][]{2003MNRAS.342..399G} and a $a_{\rm r}=-0.7$ \citep[e.g.,][]{2019A&A...630A..83Z}, this core mechanical luminosity of Hercules A is $L_{\rm mech, core}\sim 1.3-2.7 \cdot 10^{45}\, \rm erg/s$, which results to:
\begin{equation}
\lambda_{\rm curr} = \frac{L_{\rm bol,curr}+L_{\rm mech,core}}{L_{\rm Edd}} \sim 0.005 \, - 0.01.
\label{equation_curr}
\end{equation} 
Notably, this estimate is over seven times lower than the values obtained using the literature methodology. Finally, given the disproportion between the radio jet flux and core emission for Hercules A we can assume that $L_{\rm mech,past} \simeq L_{\rm mech,total}$ which for the `past' accretion disk generating the observed jets and lobes would result to:
\begin{equation}
\lambda_{\rm past} = \frac{L_{\rm bol,past}+L_{\rm mech,past}}{L_{\rm Edd}}.
\label{equation_past}
\end{equation} 

In these calculations, the absence of data on the optical emission from the past accretion disk restricts us to adopt a conservative lower limit of $L_{\rm bol,past}=0$, resulting in $\lambda_{\rm past}>0.07$. While the former results from the core emission ($\lambda_{\rm curr}$) point to a low accretion rate disk (i.e., radio mode; ADAF disk), incorporation of the extended radio emission (i.e., $\lambda_{\rm liter}$ and $\lambda_{\rm past}$) yields considerably higher accretion rates corresponding to a quasar mode, TD scenario. In this sense, Hercules A serves as a compelling example, suggesting the presence of significant systematic biases in previous studies that relied solely on the total radio flux of the radio source. Conventional state-of-the-art calculations neglect the core flux from the ongoing AGN episode, leading to inaccurate estimations of the accretion rates. Therefore, Hercules A has been categorised as a LERG \citep[e.g.,][]{2019A&A...632A.124B}, which according to the AGN unification model and our methodology is typically linked with an ADAF accretion disk. However, based on the total radio emission, previous studies would estimate a higher \textit{Eddington}-scaled accretion rate of Hercules A, positioning it as a galaxy with a quasar accretion disk mode.

\subsection{Statistical sample}
To delve deeper into this phenomenon and fully grasp its implications for unification studies and the refinement of accretion disk theories, it is imperative to conduct a more comprehensive investigation involving a larger sample size. For the application of our methodology we require, however, the knowledge of the O[III] emission line, SMBH mass and radio characteristics of radio galaxies. One catalogue providing such information for hundreds of radio galaxies is the work by \citet{2017MNRAS.466.4346M}. Their sample resulted from a cross-match of the Sloan Digital Sky Survey DR7 \citep[SDSS;][]{2009ApJS..182..543A} with the NRAO VLA Sky Survey \citep[NVSS;][]{1998AJ....115.1693C} and the Faint Images of the Radio Sky at Twenty-cm \citep[FIRST;][]{2015ApJ...801...26H} radio catalogues, with a redshift range of $0.03<z<0.3$ and a 40 mJy flux density cut. The authors of this work visually inspected and categorised these galaxies into FRI or FRII as well as into HERGs/LERGs by estimating the ratios of four high-excitation lines (O[III], N[II], S[II] and O[I]) to the H$\alpha$ and H$\beta$ emission lines. For the SMBH mass estimates they used the relation between the velocity dispersion ($\sigma$) and SMBH mass, provided by \citet{2002ApJ...574..740T}: $\textrm{log}(\rm M_{\rm SMBH}/M_{\odot}) = 8.13 + 4.02 \, \textrm{log}(\sigma/200 \, km s^{-1})$.\\ 
\indent The final sample, which consists of 137 radio galaxies (obtained by cross-matching the two sub-tables of the main catalogue) was used in conjunction with our aforementioned methodology to estimate the $\lambda_{\rm liter}$, $\lambda_{\rm curr}$ and $\lambda_{\rm past}$. To determine $L_{\rm mech,core}$, we applied a uniform radius of 10 kpc to all galaxies, removing possible extended emission potentially generated by an older accretion disk. This value is a rough estimate of a compact source and is motivated by the typical sizes observed for FR0s \citep[i.e., low-luminosity compact radio galaxies; see][]{2023A&ARv..31....3B}. In order to estimate the spectral index of each source and subsequently calculate the $L_{\rm mech}$, we employed the VLASS 3 GHz catalogue \citep[e.g.,][]{2021ApJS..255...30G} using total radio flux measurements. For the galaxies that a 3 GHz detection is reported, the distribution of spectral indices follows the expected power-law slope, with a maximum of around -0.7. For the remaining non-detected galaxies with VLASS we assume a spectral index of -0.7. For a discussion of a flatter radio core profile, the reader is directed to the Results \ref{results} section.
\begin{figure*}
\centering
\includegraphics[width=2\columnwidth]{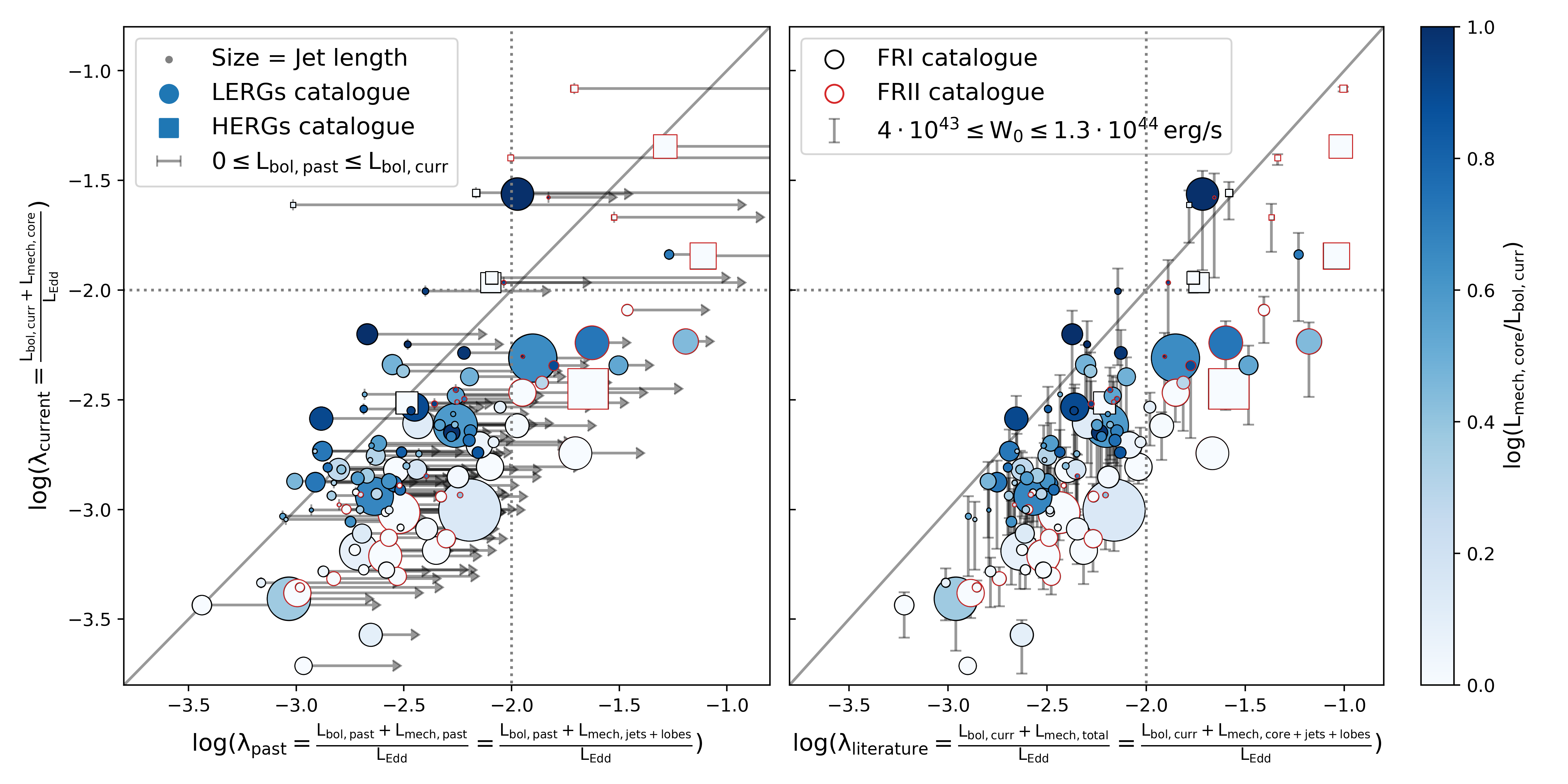}
\caption{Left panel: comparison of past (consideration of $L_{\rm 1.4GHz,jets+lobes}$; equation \ref{equation_past}) and current (using solely $L_{\rm 1.4GHz,core}$; equation \ref{equation_curr}) \textit{Eddington}-scaled accretion rates ($\lambda$) for the sample of 121 radio galaxies. Galaxies classified as LERGs and HERGs are shown in blue with circular and square markers, respectively, while black and red edges denote FRI and FRII galaxies. The intensity of each data point's colour reflects the $L_{\rm mech,core}/L_{\rm bol,curr}$ ratio (i.e., brighter red or blue colours indicate higher ratios), while the marker's size corresponds to the length of the radio jet. Error bars on the x-axis represent the lower and upper bounds of $L_{\rm bol, past}$. Right panel: comparison of the $\lambda_{\rm curr}$ with $\lambda_{\rm liter}$ (integrating the total radio flux; see equation \ref{equation0}), with the error-bars following the $W_0$ range. The dotted grey lines correspond to the accretion limit that separates quasar from radio accretion disks.}
\label{z_comparison}
\end{figure*}

While \citealt{2017MNRAS.466.4346M} have evaluated visually the morphology of each individual source using NVSS photometry, in our work, we have conducted a further visual inspection of the combined photometry from FIRST and SDSS (r-band). Our reference photometry for the radio flux estimation is FIRST, which offers the resolution required to properly identify the radio structure of each galaxy. In this manner, spurious sources for which an estimation of the core and jet radio emission could be non-trivial are excluded. Such cases include galaxies with complex radio jet morphologies and bent jets caused by interactions with the intergalactic medium. From this inspection, 16 sources were deemed unusable due to the lack of a clear radio structure. Of the remaining galaxies, 86 were identified by \citet{2017MNRAS.466.4346M} as FRI, 35 as FRII galaxies, 110 as LERGs and 11 as HERGs. Consequently, the final analysis was based on a sample of 121 radio galaxies with a redshift range of $0.05<\textrm{z}<0.1$.

\section{Results}\label{results}
The \textit{Eddington}-scaled accretion rates were computed as described above and are shown in figure \ref{z_comparison}. These include estimates of past AGN activity ($\lambda_{\rm past}$), derived from combined jet and lobe radio emission (equation \ref{equation_past}); current accretion rates ($\lambda_{\rm curr}$), based solely on the 10 kpc core radio luminosity (equation \ref{equation_curr}); and literature-based estimates ($\lambda_{\rm liter}$), following the standard approach (equations \ref{equation-1} and \ref{equation0}).
\subsection{Impact on accretion rate estimates}
In figure \ref{z_comparison}, the radio galaxies are shaped-coded based on their optical excitation line classification (i.e., LERGs or HERGs), with an additional outline indicating whether the galaxy is categorised as FRI or FRII. The colour intensity corresponds to the ratio between the core mechanical and bolometric luminosities, while the size of each point represents the maximum extent of the radio emission. Error bars on both panels of the figure depict the uncertainties related to the $L_{\rm bol,past}$ and $L_{\rm mech,core}$. This figure highlights the significant impact of our methodology on the estimation of the \textit{Eddington}-scaled accretion rate. For instance, in the past-to-current comparison (left panel), we observe a consistent overestimation of the $\lambda$ parameter ($\overline{\lambda_{\rm past}/\lambda_{\rm curr}}>2.5$), especially for the galaxies with low $L_{\rm mech,core}/L_{\rm bol,curr}$. Particularly, 13 galaxies seem to have shifted from quasar-mode accretion (i.e., $\textrm{log}\lambda_{\rm past}>-2$) to radio-mode accretion (i.e., $\textrm{log}\lambda_{\rm current}<-2$), while 5 sources exhibit the opposite shift. For the latter cases, it has to be stressed, however, that the shift is observed only at the lower limit, assuming that $L_{\rm bol,past}=0$.

\begin{figure*}
\centering
\includegraphics[width=2\columnwidth]{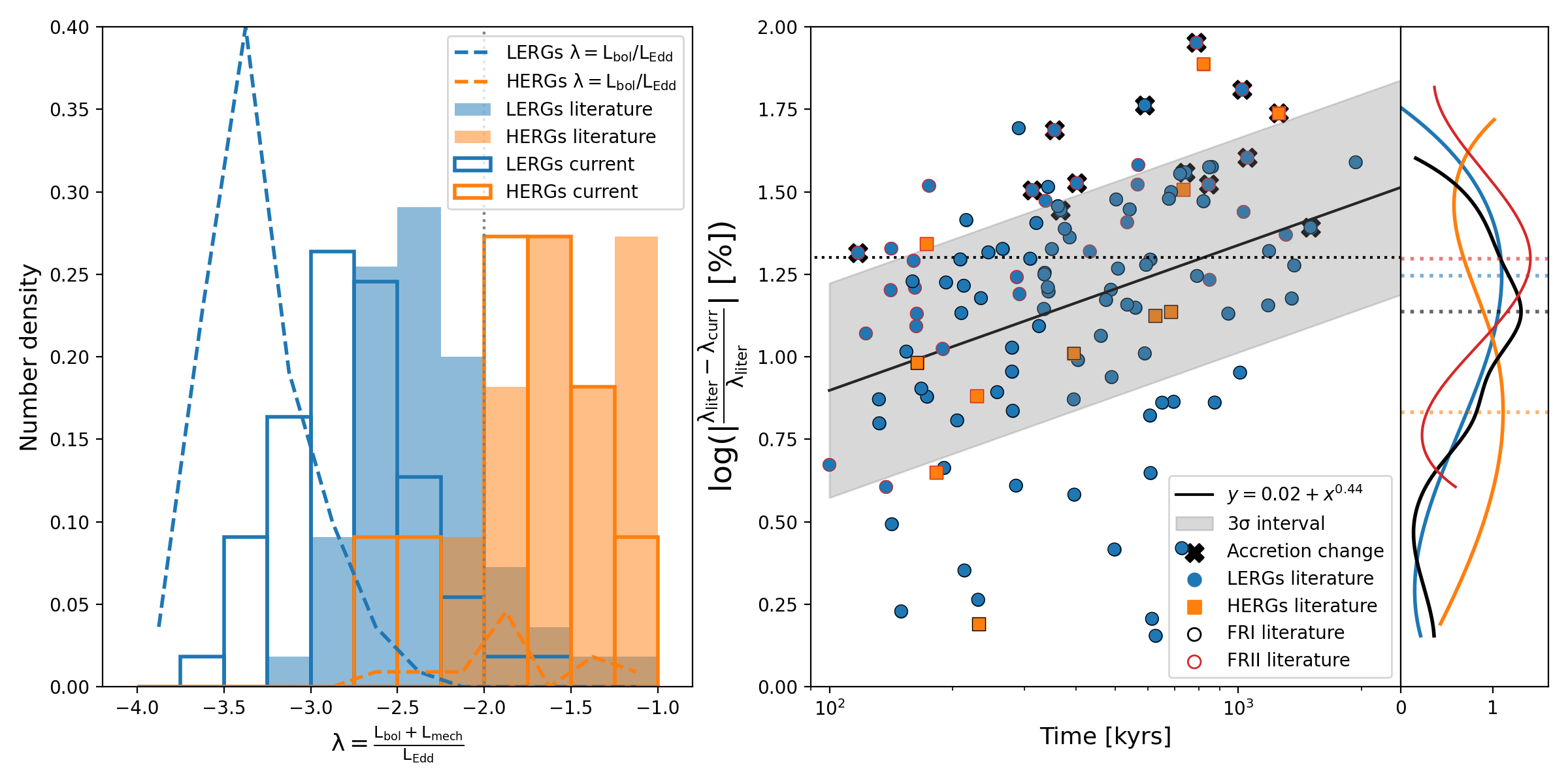}
\caption{Left panel: the number density histograms of the current $\lambda$ (empty rectangles) and literature (filled rectangles) for the LERGs and HERGs presented with blue and orange colour, respectively. The dashed lines correspond to the $\lambda$ distribution excluding the radio/mechanical luminosity (i.e., $\lambda=L_{\rm bol}/L_{\rm Edd}$), and are colored accordingly. The dotted grey line depicts the transition limit between quasar and radio accretion mode. Right panel: the difference in $\%$ between the current and literature $\lambda$ values over an estimation of the time that each jet will require to extend assuming a relativistic speed of 0.5c. Galaxies classified as LERG and HERG are shown as blue circles and orange squares, respectively, with X-shaped markers indicating sources for which their accretion rate has shifted above/below the accretion mode limit. A linear regression fit (grey shaded area) with its $3 \sigma$ interval is also included, as well as the $\Delta \lambda$ distributions for H/LERGs and FRI/IIs, annotated with their mean values and corresponding colour.}
\label{z_comparison2}
\end{figure*}

Furthermore, the current accretion rate estimates yield consistently lower $\lambda$ range of values compared to the literature ones (i.e., $\overline{\lambda_{\rm liter}/\lambda_{\rm curr}} \sim 2.6-4.7$; right panel of figure \ref{z_comparison}), with LERGs and FRII galaxies being the most affected (see also figure \ref{z_comparison2}). In a further analysis, the right panel of figure \ref{z_comparison2} illustrates the relative difference (in $\%$) between the current and literature estimates of $\lambda$, plotted against the time required for each lobe to reach its current extension (assuming a constant velocity of 0.5c; see \citealt{2006MNRAS.368..609J,2023A&A...672A.104G}). Notably, the sources most impacted (those transitioning between two accretion modes) are predominantly LERGs and FRIIs, with timescales extending up to $\sim 2 \, \rm Myrs$, suggesting that a change in accretion mode could have occurred within this period. Additionally, a correlation is observed between $\Delta \lambda = log(|(\lambda_{\rm liter}-\lambda_{\rm curr})/\lambda_{\rm liter}|)$ and the size or expansion timescale ($t$) of of the jet the radio source described with the equation $\Delta \lambda=0.22+t^{0.44}$. While this result emphasises the importance of considering only the core emission when analysing radio galaxies with extended radio jets, the observed correlation is relatively weak and subject to substantial scatter, and thus must be treated with caution. Furthermore, the choice of core radius can significantly affect the resulting differences, with a value of 10 kpc leading to an average $<\Delta \lambda>$ of approximately 26 $\%$, whereas a radius of 30 kpc yields a much smaller difference of about 3.4 $\%$.

Moreover, the left panel of figure \ref{z_comparison2} displays the distribution of the $\lambda$ values for the current and literature estimates, with blue and orange histograms representing the LERG and HERG classification, respectively. Additionally, for reference, the accretion rates $\lambda = L_{\rm bol}/L_{\rm Edd}$, excluding the radio contribution, are also included. We note that these LERG and HERG distributions fall within the ranges typically expected for such populations, in agreement with previous findings \citep[e.g.,][]{2012MNRAS.421.1569B}. Using our revised, current $\lambda$ estimates, we find that LERGs are more likely to exhibit radio-mode accretion (i.e., $log \lambda < -2$), with just 3 $\%$ falling in the quasar mode regime (i.e., $log \lambda > -2$). In contrast, the literature estimates present a less distinct relationship, with 13$\%$ of LERGs included in the quasar accretion mode regime. For HERGs, due to their smaller and more concentrated radio emission, the difference between current and literature $\lambda$ values is negligible. 

In conclusion, our analysis suggests that, using this sample of 121 radio galaxies, literature studies would have estimated a different accretion mode for 10 $\%$ of LERGs and 1 $\%$ of HERGs with respect to their `true', current accretion rate modes calculated using our method. If this effect holds across redshift, then in studies such as \citet{2025MNRAS.539..463C}, where 2694 out of 2828 radio galaxies were classified as LERGs, approximately 269 may have inaccurately inferred accretion rates. It is important to emphasize that the spectral index used in equation \ref{equation3} is not the flat core value reported in studies such as \citet{2007MNRAS.381..589M}, but rather the index derived from the total source flux density using VLASS data at 3 GHz (or -0.7 where data were missing). Nevertheless, the effect of assuming a steeper core profile is minimal, with a flatter index still resulting in only 10 $\%$ and 0.8 $\%$ of sources being affected for the LERGs and HERGs, respectively. Such a decision is motivated by the fact that, as it has been documented in the literature \citep[e.g.,][]{2022PASA...39...21M}, many galaxies seem to have steeper radio core profiles, such as Hercules A.

\subsection{Caveats}\label{caveats}
Several important caveats apply to our analysis, including the use of equation \ref{equation3}, the adopted SMBH masses, and other underlying assumptions, which we discuss in detail below.

Primarily, a key limitation of our methodology lies in the use of the normalisation parameter $W_{\rm 0}$ in equation \ref{equation3}, which is typically derived in the literature based on X-ray cavities inflated by extended radio jets. In this work, we consider that these cavities are relics of past AGN activity and do not directly trace the current state of the accretion disk. Our approach aims to estimate the present-day \textit{Eddington}-scaled accretion rate of SMBHs under the assumption that it reflects the kinetic output of the jets currently observed as core radio emission. This connection is inherently uncertain from an observational perspective, as the evolution of radio jets cannot be reliably predicted. Therefore, we adopt this normalisation range due to these observational constraints, its coverage of a broad spectrum of jet power, and its theoretical grounding in high-resolution jet simulations \citep[e.g.,][]{2023A&A...679A.160O}. Regarding the latter, such studies follow three-dimensional, magnetohydrodynamic simulations of sub-relativistic jets, demonstrating that the relationship between simulated cavity power and radio luminosity (see their figure 12) is broadly consistent with observations, thereby supporting the empirical relations adopted in our work.

\begin{figure}
\centering
\includegraphics[width=\columnwidth]{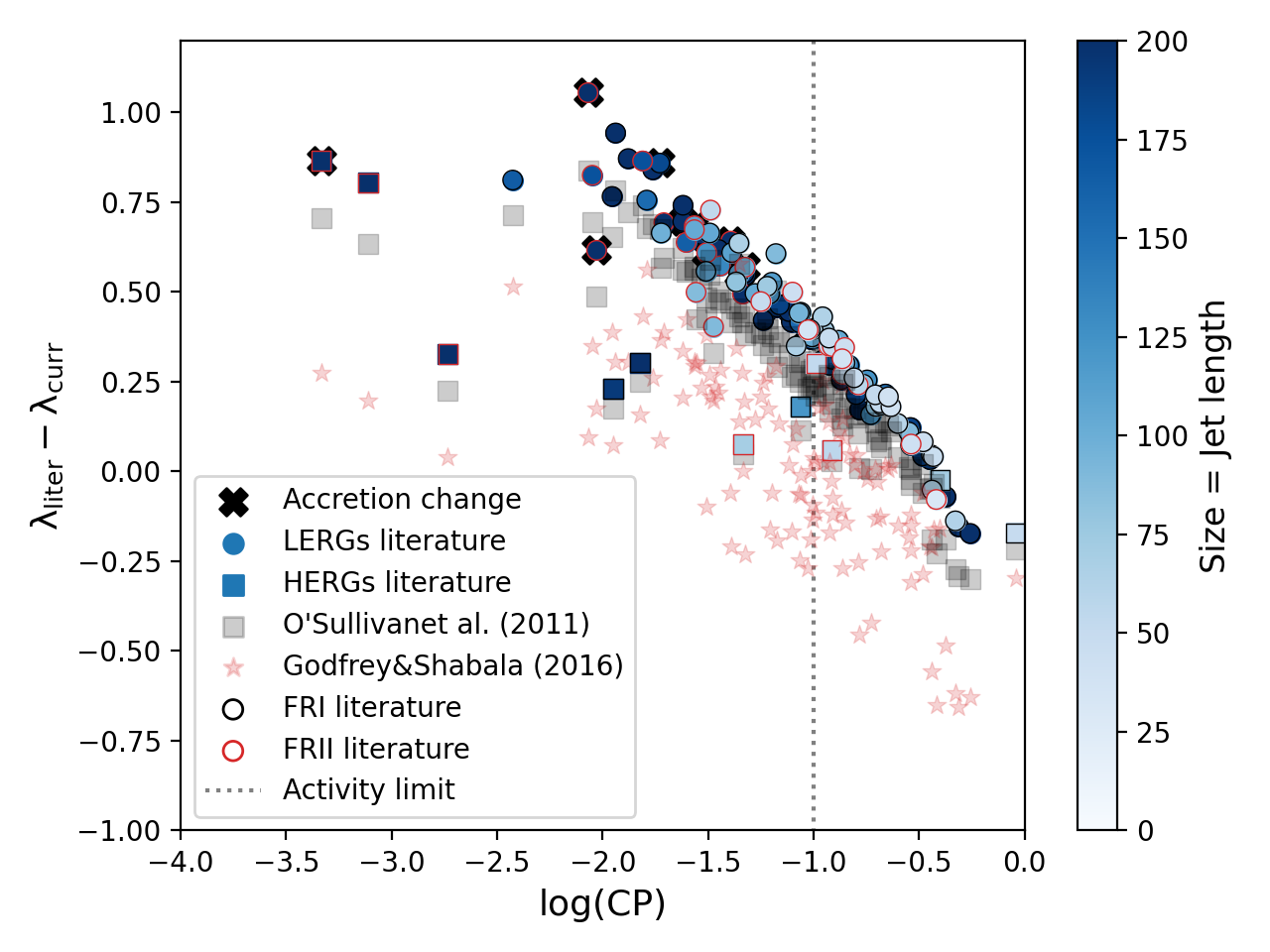}
\caption{The absolute difference between $\lambda_{\mathrm{liter}}$ and $\lambda_{\mathrm{curr}}$ as a function of the Core Prominence (CP) for our sample. The colour intensity of each point represents the jet length of the corresponding radio galaxy, while the markers layout follows the same scheme as in figure \ref{z_comparison}. The dotted grey vertical line indicates the threshold for the restarted AGN phase (CP = 0.1), as proposed by \citet{2020A&A...638A..34J}.}
\label{lambda_cp}
\end{figure}

To further explore the implications of our methodology, we examine the Core Prominence (CP) defined as the ratio of core to total radio flux \citep{2024A&A...691A.287N}, in relation to other physical properties of the galaxies in our sample. Figure \ref{lambda_cp} presents the results, showing the absolute difference between the \textit{Eddington}-scaled accretion rates ($\lambda$) from the literature and those derived using our method, plotted against the CP of each source. From these results, we can conclude that our sample primarily consists of typical radio-loud galaxies with low CP (median value of 0.08), consistent with them not being in a restarted AGN phase, according to the criteria by literature studies \citep[CP<0.1-0.2; e.g.][]{2020A&A...638A..34J}. 

In the context of radio galaxy evolution, the elevated CP values (i.e., CP>0.1-0.2) observed in restarted sources can be interpreted as indicative of the jet power during the initial active phase. In comparison, the majority of extended radio sources exhibit significantly lower CP values, commonly in the range of 0.01 to 0.05 (see figure \ref{lambda_cp}). Assuming a direct correlation between CP and jet power \citep[e.g.,][]{2025MNRAS.539..463C}, this suggests that the average jet power in these extended sources has decreased by approximately an order of magnitude relative to its original value. Consequently, the CP of restarted sources may serve as a practical reference point for the jet power of young sources, offering a useful basis to quantify the typical decline in jet activity over the lifetime of radio galaxies.

Despite observational uncertainties, our results indicate that galaxies with the lowest CP (i.e, potential candidates for being in a restarted or remnant phase) exhibit the largest discrepancies between our accretion rate estimates and those reported in previous studies. Considering that empirical relations in the literature connecting mechanical output and radio power show significant variation \citep[see for instance][]{1999MNRAS.309.1017W,2017MNRAS.467.1586I,smolcic17} mainly due to the limited amount of available data, the values of $\lambda_{\mathrm{liter}}$ can also vary and thus affect our results. To address this, we included in figure \ref{lambda_cp} the $\Delta \lambda$ values obtained using two additional empirical relations \citep[by][the latter incorporates a luminosity-distance–adjusted estimation]{2011ApJ...735...11O,2016MNRAS.456.1172G}. From this comparison, we observe a systematic offset in $\Delta \lambda$ relative to our original calculation, of -0.08 and -0.27 for the respective works. Interestingly, the calculations based on the empirical relation by \citet{2011ApJ...735...11O} yield results consistent with our original estimates, with an average $\Delta \lambda$ of $\sim 0.3$. In contrast, the relation from \citet{2016MNRAS.456.1172G} indicates that accretion rates are systematically underestimated in contrast with our results, highlighting the need for caution when applying such empirical relations. 

This implies that traditional methods, which rely on total radio emission, may overestimate accretion rates in such sources by failing to isolate the current nuclear activity traced by the core. In this sense, equation \ref{equation3}, combined with the adopted $W_{\rm 0}$ range, might offer a more refined view of SMBH accretion, potentially correcting for biases in earlier estimates. It should be emphasised, however, that this correlation is partly driven by the fact that galaxies with lower CP values typically exhibit lower radio core luminosities, resulting in larger $\Delta \lambda$ differences. Correlations between CP and other physical properties, such as bolometric luminosity and spectral index, were also examined, generally revealing only weak or marginal trends. However, a correlation between bolometric luminosity and CP was identified among FRII sources (\textit{Kendall}-test $\tau$ value of -0.3 and p-value of 0.01) following the relation: $log(L_{\rm bol})=39.47 -0.3 CP$.

In addition to the challenges of estimating $L_{\rm bol,past}$ and $L_{\rm mech,core}$, further uncertainties arise when inferring past accretion disk properties from current SMBH measurements, as well as from potential contamination of the radio luminosity by star formation rate (SFR). However, we argue that both effects have a negligible impact on our estimates. Considering that even the most extended radio jets of our sample have expanded to approximately 300 kpc, it would require less than roughly 10 Myr to propagate to such a distance. Consequently, we can reasonably assume that the SMBH masses employed from the catalogue (with a minimum value of $10^7 \, \rm M_{\odot}$), would not undergo significant increase even under super-\textit{Eddington} (SE) accretion rates (a $\sim 10\%$ increase in mass is expected at SE accretion for $M_{\rm SMBH}\sim 10^{8} \, \rm M_{\odot}$). Nonetheless, it has to be noted that the SMBH mass estimates of our sample are derived from an empirical relation with large uncertainties, which might be impacted by the fact that the bulge formation and SMBH evolution are tightly linked. Such scatter could potentially affect the classification of the accretion rate regime, especially for the sources that are close to the 1 $\%$ limit of \textit{Eddington} accretion rate. Similarly, the scatter in the empirical relation of equation \ref{equation-1} could also impact the results in a similar manner, nevertheless, this effect would impact solely the total mechanical output rather than the core emission.\\
\indent Another potential complication arises from high SFR, which can primarily affect the 1.4 GHz radio luminosities and further impact the estimation of the $L_{\rm [OIII]}$ \citep[see][for a review]{heckman14}. While a spectral energy distribution (SED) fitting approach could help disentangle these effects, we opted against this analysis due to the absence of multi-wavelength data and sufficient resolution for many sources. Additionally, unresolved sources may indicate systems of interacting galaxies, where one galaxy drives the SFR and another powers the radio jet \citep[as detailed in][]{2023A&A...678A.116A}, which might complicate even more such analysis. Importantly, this issue might impact only the comparison between $\lambda_{\rm past}$ and $\lambda_{\rm curr}$, as both $\lambda_{\rm curr}$ and $\lambda_{\rm liter}$ rely on current bolometric luminosity estimates and are similarly affected. The comparison would only be compromised if the bolometric luminosity significantly exceeded the total radio luminosity, which is uncommon for FR-bright radio sources. In our sample, the average bolometric luminosity is $\overline{\textrm{log}L_{\rm bol,curr}} \sim 43.33 \pm 0.42 \, \rm erg/s$, while the average total mechanical luminosity is $\overline{\textrm{log}L_{\rm mech,total}} \sim 44.16 \pm 0.32 \, \rm erg/s$. Furthermore, \citet{2012MNRAS.421.1569B} identified, via the Baldwin, Phillips $\&$ Terlevich methodology \citep[BPT;][]{1981PASP...93....5B} all galaxies in this sample as radio-loud AGN (with $L_{\rm 1.4GHz}>10^{24} \, \rm W/Hz$) rather than star-forming galaxies, supporting the current classification of the sources and minimising the potential impact of SFR on our results.

\section{Conclusions}
With this brief study, we would like to instigate a discussion regarding the importance of distinguishing between the core and jet/lobe radio emission of radio-loud galaxies and their optical core emission when estimating the \textit{Eddington}-scaled accretion rate. Our findings suggest that existing literature may overestimate this accretion rate, leading to the classification of some radio galaxies into different accretion modes. This discrepancy is particularly pronounced in the LERG and FRII populations of our sample and could have significant implications for high-redshift radio galaxies, where resolution limitations hinder the clear identification of radio jets. Recent advances in radio interferometry have led to the discovery of galaxies at high-z, with resolved jets extending up to $\sim 1$ kpc \citep[e.g.,][]{2018ApJ...861...86M}. Despite the short emission timescales of 5-20 kyr, our methodology can prove useful in studying episodic accretion events, which are frequent during these early cosmic epochs. For instance, just as some low-redshift radio galaxies of our sample are classified differently due to their accretion rate regimes, high-redshift radio-loud galaxies with kiloparsec-scale jet extensions and SMBH masses of the order of $M_{\rm SMBH}\sim 10^8 \, \rm M_{\odot}$ \citep[e.g.,][]{2025ApJ...980L...8G} could be similarly affected. It is important to note that the empirical relations employed in this work are derived from low-$z$ radio galaxies and should be applied with caution when dealing with the high-redshift Universe. In addition, extended features of radio galaxies, such as jets, may remain unresolved at high redshifts or at MHz frequencies, further limiting the applicability of such methodology. With high-resolution data, however, our approach can be applied to estimate accretion rate correction factors, and future work could aim to derive these systematically. Such corrections may also have substantial implications for studies that infer the spin parameter of SMBHs from radio emission, underscoring the importance of incorporating these considerations in future research.

\section*{Acknowledgements}
I.B. has received funding from the European Union's Horizon 2020 research and innovation programme under the Marie Sklodowska-Curie Grant agreement ID n.º 101059532. We would like thank the anonymous referee for their valuable comments which have significantly enhanced the quality and clarity of this article.

\section*{Data Availability}
The data underlying this article will be shared on reasonable request to the corresponding author. The datasets were derived from sources in the public domain: \url{https://cdsarc.cds.unistra.fr/viz-bin/cat/J/MNRAS/466/4346}.



\bibliographystyle{mnras}
\bibliography{main} 

\begin{thebibliography}{}
\makeatletter
\relax
\def\mn@urlcharsother{\let\do\@makeother \do\$\do\&\do\#\do\^\do\_\do\%\do\~}
\def\mn@doi{\begingroup\mn@urlcharsother \@ifnextchar [ {\mn@doi@}
  {\mn@doi@[]}}
\def\mn@doi@[#1]#2{\def\@tempa{#1}\ifx\@tempa\@empty \href
  {http://dx.doi.org/#2} {doi:#2}\else \href {http://dx.doi.org/#2} {#1}\fi
  \endgroup}
\def\mn@eprint#1#2{\mn@eprint@#1:#2::\@nil}
\def\mn@eprint@arXiv#1{\href {http://arxiv.org/abs/#1} {{\tt arXiv:#1}}}
\def\mn@eprint@dblp#1{\href {http://dblp.uni-trier.de/rec/bibtex/#1.xml}
  {dblp:#1}}
\def\mn@eprint@#1:#2:#3:#4\@nil{\def\@tempa {#1}\def\@tempb {#2}\def\@tempc
  {#3}\ifx \@tempc \@empty \let \@tempc \@tempb \let \@tempb \@tempa \fi \ifx
  \@tempb \@empty \def\@tempb {arXiv}\fi \@ifundefined
  {mn@eprint@\@tempb}{\@tempb:\@tempc}{\expandafter \expandafter \csname
  mn@eprint@\@tempb\endcsname \expandafter{\@tempc}}}

\bibitem[\protect\citeauthoryear{{Abazajian} et~al.,}{{Abazajian}
  et~al.}{2009}]{2009ApJS..182..543A}
{Abazajian} K.~N.,  et~al., 2009, \mn@doi [\apjs]
  {10.1088/0067-0049/182/2/543}, \href
  {https://ui.adsabs.harvard.edu/abs/2009ApJS..182..543A} {182, 543}

\bibitem[\protect\citeauthoryear{{Amarantidis}, {Afonso}, {Matute}, {Farrah},
  {Hopkins}, {Messias}, {Pappalardo}  \& {Seymour}}{{Amarantidis}
  et~al.}{2023}]{2023A&A...678A.116A}
{Amarantidis} S.,  {Afonso} J.,  {Matute} I.,  {Farrah} D.,  {Hopkins} A.,
  {Messias} H.,  {Pappalardo} C.,   {Seymour} N.,  2023, \mn@doi [\aap]
  {10.1051/0004-6361/202346411}, \href
  {https://ui.adsabs.harvard.edu/abs/2023A&A...678A.116A} {678, A116}

\bibitem[\protect\citeauthoryear{{Baldi}}{{Baldi}}{2023}]{2023A&ARv..31....3B}
{Baldi} R.~D.,  2023, \mn@doi [\aapr] {10.1007/s00159-023-00148-3}, \href
  {https://ui.adsabs.harvard.edu/abs/2023A&ARv..31....3B} {31, 3}

\bibitem[\protect\citeauthoryear{{Baldwin}, {Phillips}  \&
  {Terlevich}}{{Baldwin} et~al.}{1981}]{1981PASP...93....5B}
{Baldwin} J.~A.,  {Phillips} M.~M.,   {Terlevich} R.,  1981, \mn@doi [\pasp]
  {10.1086/130766}, \href
  {https://ui.adsabs.harvard.edu/abs/1981PASP...93....5B} {93, 5}

\bibitem[\protect\citeauthoryear{{Balmaverde} et~al.,}{{Balmaverde}
  et~al.}{2019}]{2019A&A...632A.124B}
{Balmaverde} B.,  et~al., 2019, \mn@doi [\aap] {10.1051/0004-6361/201935544},
  \href {https://ui.adsabs.harvard.edu/abs/2019A&A...632A.124B} {632, A124}

\bibitem[\protect\citeauthoryear{{Balmaverde} et~al.,}{{Balmaverde}
  et~al.}{2021}]{2021A&A...645A..12B}
{Balmaverde} B.,  et~al., 2021, \mn@doi [\aap] {10.1051/0004-6361/202039062},
  \href {https://ui.adsabs.harvard.edu/abs/2021A&A...645A..12B} {645, A12}

\bibitem[\protect\citeauthoryear{{Best} \& {Heckman}}{{Best} \&
  {Heckman}}{2012}]{2012MNRAS.421.1569B}
{Best} P.~N.,  {Heckman} T.~M.,  2012, \mn@doi [\mnras]
  {10.1111/j.1365-2966.2012.20414.x}, \href
  {https://ui.adsabs.harvard.edu/abs/2012MNRAS.421.1569B} {421, 1569}

\bibitem[\protect\citeauthoryear{{Cavagnolo}, {McNamara}, {Nulsen}, {Carilli},
  {Jones}  \& {B{\^\i}rzan}}{{Cavagnolo} et~al.}{2010}]{2010ApJ...720.1066C}
{Cavagnolo} K.~W.,  {McNamara} B.~R.,  {Nulsen} P.~E.~J.,  {Carilli} C.~L.,
  {Jones} C.,   {B{\^\i}rzan} L.,  2010, \mn@doi [\apj]
  {10.1088/0004-637X/720/2/1066}, \href
  {https://ui.adsabs.harvard.edu/abs/2010ApJ...720.1066C} {720, 1066}

\bibitem[\protect\citeauthoryear{{Chilufya}, {Hardcastle}, {Pierce}, {Drake},
  {Baldi}, {R{\"o}ttgering}  \& {Smith}}{{Chilufya}
  et~al.}{2025}]{2025MNRAS.539..463C}
{Chilufya} J.,  {Hardcastle} M.~J.,  {Pierce} J.~C.~S.,  {Drake} A.~B.,
  {Baldi} R.~D.,  {R{\"o}ttgering} H.~J.~A.,   {Smith} D.~J.~B.,  2025, \mn@doi
  [\mnras] {10.1093/mnras/staf508}, \href
  {https://ui.adsabs.harvard.edu/abs/2025MNRAS.539..463C} {539, 463}

\bibitem[\protect\citeauthoryear{{Condon}, {Cotton}, {Greisen}, {Yin},
  {Perley}, {Taylor}  \& {Broderick}}{{Condon}
  et~al.}{1998}]{1998AJ....115.1693C}
{Condon} J.~J.,  {Cotton} W.~D.,  {Greisen} E.~W.,  {Yin} Q.~F.,  {Perley}
  R.~A.,  {Taylor} G.~B.,   {Broderick} J.~J.,  1998, \mn@doi [\apj]
  {10.1086/300337}, \href
  {https://ui.adsabs.harvard.edu/abs/1998AJ....115.1693C} {115, 1693}

\bibitem[\protect\citeauthoryear{{Fabian}}{{Fabian}}{2012}]{fabian12}
{Fabian} A.~C.,  2012, \mn@doi [\ARAA] {10.1146/annurev-astro-081811-125521},
  \href {http://adsabs.harvard.edu/abs/2012ARA%26A..50..455F} {50, 455}

\bibitem[\protect\citeauthoryear{{Fanaroff} \& {Riley}}{{Fanaroff} \&
  {Riley}}{1974}]{1974MNRAS.167P..31F}
{Fanaroff} B.~L.,  {Riley} J.~M.,  1974, \mn@doi [\mnras]
  {10.1093/mnras/167.1.31P}, \href
  {https://ui.adsabs.harvard.edu/abs/1974MNRAS.167P..31F} {167, 31P}

\bibitem[\protect\citeauthoryear{{Fujita}, {Kawakatu}  \& {Shlosman}}{{Fujita}
  et~al.}{2016}]{2016PASJ...68...26F}
{Fujita} Y.,  {Kawakatu} N.,   {Shlosman} I.,  2016, \mn@doi [Publications of
  the Astronomical Society of Japan] {10.1093/pasj/psw012}, \href
  {https://ui.adsabs.harvard.edu/abs/2016PASJ...68...26F} {68, 26}

\bibitem[\protect\citeauthoryear{{Giovannini}, {Baldi}, {Capetti}, {Giroletti}
  \& {Lico}}{{Giovannini} et~al.}{2023}]{2023A&A...672A.104G}
{Giovannini} G.,  {Baldi} R.~D.,  {Capetti} A.,  {Giroletti} M.,   {Lico} R.,
  2023, \mn@doi [\aap] {10.1051/0004-6361/202245395}, \href
  {https://ui.adsabs.harvard.edu/abs/2023A&A...672A.104G} {672, A104}

\bibitem[\protect\citeauthoryear{{Gizani} \& {Leahy}}{{Gizani} \&
  {Leahy}}{2003}]{2003MNRAS.342..399G}
{Gizani} N. A.~B.,  {Leahy} J.~P.,  2003, \mn@doi [\mnras]
  {10.1046/j.1365-8711.2003.06469.x}, \href
  {https://ui.adsabs.harvard.edu/abs/2003MNRAS.342..399G} {342, 399}

\bibitem[\protect\citeauthoryear{{Gloudemans} et~al.,}{{Gloudemans}
  et~al.}{2025}]{2025ApJ...980L...8G}
{Gloudemans} A.~J.,  et~al., 2025, \mn@doi [\apjl] {10.3847/2041-8213/ad9609},
  \href {https://ui.adsabs.harvard.edu/abs/2025ApJ...980L...8G} {980, L8}

\bibitem[\protect\citeauthoryear{{Godfrey} \& {Shabala}}{{Godfrey} \&
  {Shabala}}{2016}]{2016MNRAS.456.1172G}
{Godfrey} L.~E.~H.,  {Shabala} S.~S.,  2016, \mn@doi [\mnras]
  {10.1093/mnras/stv2712}, \href
  {https://ui.adsabs.harvard.edu/abs/2016MNRAS.456.1172G} {456, 1172}

\bibitem[\protect\citeauthoryear{{Gordon} et~al.,}{{Gordon}
  et~al.}{2021}]{2021ApJS..255...30G}
{Gordon} Y.~A.,  et~al., 2021, \mn@doi [\apjs] {10.3847/1538-4365/ac05c0},
  \href {https://ui.adsabs.harvard.edu/abs/2021ApJS..255...30G} {255, 30}

\bibitem[\protect\citeauthoryear{{Grandi}, {Torresi}, {Macconi}, {Boccardi}  \&
  {Capetti}}{{Grandi} et~al.}{2021}]{2021ApJ...911...17G}
{Grandi} P.,  {Torresi} E.,  {Macconi} D.,  {Boccardi} B.,   {Capetti} A.,
  2021, \mn@doi [\apj] {10.3847/1538-4357/abe776}, \href
  {https://ui.adsabs.harvard.edu/abs/2021ApJ...911...17G} {911, 17}

\bibitem[\protect\citeauthoryear{{Hardcastle} \& {Croston}}{{Hardcastle} \&
  {Croston}}{2020}]{2020NewAR..8801539H}
{Hardcastle} M.~J.,  {Croston} J.~H.,  2020, \mn@doi [New A]
  {10.1016/j.newar.2020.101539}, \href
  {https://ui.adsabs.harvard.edu/abs/2020NewAR..8801539H} {88, 101539}

\bibitem[\protect\citeauthoryear{{Heckman} \& {Best}}{{Heckman} \&
  {Best}}{2014}]{heckman14}
{Heckman} T.~M.,  {Best} P.~N.,  2014, \mn@doi [\ARAA]
  {10.1146/annurev-astro-081913-035722}, \href
  {http://adsabs.harvard.edu/abs/2014ARA%26A..52..589H} {52, 589}

\bibitem[\protect\citeauthoryear{{Heinz}, {Merloni}  \& {Schwab}}{{Heinz}
  et~al.}{2007}]{2007ApJ...658L...9H}
{Heinz} S.,  {Merloni} A.,   {Schwab} J.,  2007, \mn@doi [\apjl]
  {10.1086/513507}, \href
  {https://ui.adsabs.harvard.edu/abs/2007ApJ...658L...9H} {658, L9}

\bibitem[\protect\citeauthoryear{{Heinzeller} \& {Duschl}}{{Heinzeller} \&
  {Duschl}}{2007}]{2007MNRAS.374.1146H}
{Heinzeller} D.,  {Duschl} W.~J.,  2007, \mn@doi [\mnras]
  {10.1111/j.1365-2966.2006.11233.x}, \href
  {https://ui.adsabs.harvard.edu/abs/2007MNRAS.374.1146H} {374, 1146}

\bibitem[\protect\citeauthoryear{{Helfand}, {White}  \& {Becker}}{{Helfand}
  et~al.}{2015}]{2015ApJ...801...26H}
{Helfand} D.~J.,  {White} R.~L.,   {Becker} R.~H.,  2015, \mn@doi [\apj]
  {10.1088/0004-637X/801/1/26}, \href
  {https://ui.adsabs.harvard.edu/abs/2015ApJ...801...26H} {801, 26}

\bibitem[\protect\citeauthoryear{{Ineson}, {Croston}, {Hardcastle}  \&
  {Mingo}}{{Ineson} et~al.}{2017}]{2017MNRAS.467.1586I}
{Ineson} J.,  {Croston} J.~H.,  {Hardcastle} M.~J.,   {Mingo} B.,  2017,
  \mn@doi [\mnras] {10.1093/mnras/stx189}, \href
  {https://ui.adsabs.harvard.edu/abs/2017MNRAS.467.1586I} {467, 1586}

\bibitem[\protect\citeauthoryear{{Jetha}, {Hardcastle}  \& {Sakelliou}}{{Jetha}
  et~al.}{2006}]{2006MNRAS.368..609J}
{Jetha} N.~N.,  {Hardcastle} M.~J.,   {Sakelliou} I.,  2006, \mn@doi [\mnras]
  {10.1111/j.1365-2966.2006.10155.x}, \href
  {https://ui.adsabs.harvard.edu/abs/2006MNRAS.368..609J} {368, 609}

\bibitem[\protect\citeauthoryear{{Jurlin} et~al.,}{{Jurlin}
  et~al.}{2020}]{2020A&A...638A..34J}
{Jurlin} N.,  et~al., 2020, \mn@doi [\aap] {10.1051/0004-6361/201936955}, \href
  {https://ui.adsabs.harvard.edu/abs/2020A&A...638A..34J} {638, A34}

\bibitem[\protect\citeauthoryear{{Kondapally} et~al.,}{{Kondapally}
  et~al.}{2022}]{2022MNRAS.513.3742K}
{Kondapally} R.,  et~al., 2022, \mn@doi [\mnras] {10.1093/mnras/stac1128},
  \href {https://ui.adsabs.harvard.edu/abs/2022MNRAS.513.3742K} {513, 3742}

\bibitem[\protect\citeauthoryear{{Laing}, {Jenkins}, {Wall}  \&
  {Unger}}{{Laing} et~al.}{1994}]{1994ASPC...54..201L}
{Laing} R.~A.,  {Jenkins} C.~R.,  {Wall} J.~V.,   {Unger} S.~W.,  1994, in
  {Bicknell} G.~V.,  {Dopita} M.~A.,   {Quinn} P.~J.,  eds,  Astronomical
  Society of the Pacific Conference Series Vol. 54, The Physics of Active
  Galaxies. p.~201

\bibitem[\protect\citeauthoryear{{Mazoochi}, {Miraghaei}  \&
  {Riazi}}{{Mazoochi} et~al.}{2022}]{2022PASA...39...21M}
{Mazoochi} F.,  {Miraghaei} H.,   {Riazi} N.,  2022, \mn@doi [\pasa]
  {10.1017/pasa.2022.15}, \href
  {https://ui.adsabs.harvard.edu/abs/2022PASA...39...21M} {39, e021}

\bibitem[\protect\citeauthoryear{{Merloni} \& {Heinz}}{{Merloni} \&
  {Heinz}}{2007}]{2007MNRAS.381..589M}
{Merloni} A.,  {Heinz} S.,  2007, \mn@doi [\mnras]
  {10.1111/j.1365-2966.2007.12253.x}, \href
  {https://ui.adsabs.harvard.edu/abs/2007MNRAS.381..589M} {381, 589}

\bibitem[\protect\citeauthoryear{{Mingo} et~al.,}{{Mingo}
  et~al.}{2022}]{2022MNRAS.511.3250M}
{Mingo} B.,  et~al., 2022, \mn@doi [\mnras] {10.1093/mnras/stac140}, \href
  {https://ui.adsabs.harvard.edu/abs/2022MNRAS.511.3250M} {511, 3250}

\bibitem[\protect\citeauthoryear{{Miraghaei} \& {Best}}{{Miraghaei} \&
  {Best}}{2017}]{2017MNRAS.466.4346M}
{Miraghaei} H.,  {Best} P.~N.,  2017, \mn@doi [\mnras] {10.1093/mnras/stx007},
  \href {https://ui.adsabs.harvard.edu/abs/2017MNRAS.466.4346M} {466, 4346}

\bibitem[\protect\citeauthoryear{{Momjian}, {Carilli}, {Ba{\~n}ados}, {Walter}
  \& {Venemans}}{{Momjian} et~al.}{2018}]{2018ApJ...861...86M}
{Momjian} E.,  {Carilli} C.~L.,  {Ba{\~n}ados} E.,  {Walter} F.,   {Venemans}
  B.~P.,  2018, \mn@doi [\apj] {10.3847/1538-4357/aac76f}, \href
  {https://ui.adsabs.harvard.edu/abs/2018ApJ...861...86M} {861, 86}

\bibitem[\protect\citeauthoryear{{Moravec}, {Svoboda}, {Borkar}, {Boorman},
  {Kynoch}, {Panessa}, {Mingo}  \& {Guainazzi}}{{Moravec}
  et~al.}{2022}]{2022A&A...662A..28M}
{Moravec} E.,  {Svoboda} J.,  {Borkar} A.,  {Boorman} P.,  {Kynoch} D.,
  {Panessa} F.,  {Mingo} B.,   {Guainazzi} M.,  2022, \mn@doi [\aap]
  {10.1051/0004-6361/202142870}, \href
  {https://ui.adsabs.harvard.edu/abs/2022A&A...662A..28M} {662, A28}

\bibitem[\protect\citeauthoryear{{Nair} et~al.,}{{Nair}
  et~al.}{2024}]{2024A&A...691A.287N}
{Nair} D.~G.,  et~al., 2024, \mn@doi [\aap] {10.1051/0004-6361/202451522},
  \href {https://ui.adsabs.harvard.edu/abs/2024A&A...691A.287N} {691, A287}

\bibitem[\protect\citeauthoryear{{Novak}, {Ostriker}  \& {Ciotti}}{{Novak}
  et~al.}{2011}]{2011ApJ...737...26N}
{Novak} G.~S.,  {Ostriker} J.~P.,   {Ciotti} L.,  2011, \mn@doi [\apj]
  {10.1088/0004-637X/737/1/26}, \href
  {https://ui.adsabs.harvard.edu/abs/2011ApJ...737...26N} {737, 26}

\bibitem[\protect\citeauthoryear{{O'Dea} \& {Saikia}}{{O'Dea} \&
  {Saikia}}{2021}]{2021A&ARv..29....3O}
{O'Dea} C.~P.,  {Saikia} D.~J.,  2021, \mn@doi [\aapr]
  {10.1007/s00159-021-00131-w}, \href
  {https://ui.adsabs.harvard.edu/abs/2021A&ARv..29....3O} {29, 3}

\bibitem[\protect\citeauthoryear{{O'Sullivan}, {Giacintucci}, {David}, {Gitti},
  {Vrtilek}, {Raychaudhury}  \& {Ponman}}{{O'Sullivan}
  et~al.}{2011}]{2011ApJ...735...11O}
{O'Sullivan} E.,  {Giacintucci} S.,  {David} L.~P.,  {Gitti} M.,  {Vrtilek}
  J.~M.,  {Raychaudhury} S.,   {Ponman} T.~J.,  2011, \mn@doi [\apj]
  {10.1088/0004-637X/735/1/11}, \href
  {https://ui.adsabs.harvard.edu/abs/2011ApJ...735...11O} {735, 11}

\bibitem[\protect\citeauthoryear{{Ohmura} \& {Machida}}{{Ohmura} \&
  {Machida}}{2023}]{2023A&A...679A.160O}
{Ohmura} T.,  {Machida} M.,  2023, \mn@doi [\aap]
  {10.1051/0004-6361/202244690}, \href
  {https://ui.adsabs.harvard.edu/abs/2023A&A...679A.160O} {679, A160}

\bibitem[\protect\citeauthoryear{{Padovani} et~al.,}{{Padovani}
  et~al.}{2017}]{2017A&ARv..25....2P}
{Padovani} P.,  et~al., 2017, \mn@doi [The Astronomy and Astrophysics Review]
  {10.1007/s00159-017-0102-9}, \href
  {https://ui.adsabs.harvard.edu/abs/2017A&ARv..25....2P} {25, 2}

\bibitem[\protect\citeauthoryear{{Rafferty}, {McNamara}, {Nulsen}  \&
  {Wise}}{{Rafferty} et~al.}{2006}]{2006ApJ...652..216R}
{Rafferty} D.~A.,  {McNamara} B.~R.,  {Nulsen} P.~E.~J.,   {Wise} M.~W.,  2006,
  \mn@doi [\apj] {10.1086/507672}, \href
  {https://ui.adsabs.harvard.edu/abs/2006ApJ...652..216R} {652, 216}

\bibitem[\protect\citeauthoryear{{Raj} \& {Nixon}}{{Raj} \&
  {Nixon}}{2021}]{2021ApJ...909...82R}
{Raj} A.,  {Nixon} C.~J.,  2021, \mn@doi [\apj] {10.3847/1538-4357/abdc25},
  \href {https://ui.adsabs.harvard.edu/abs/2021ApJ...909...82R} {909, 82}

\bibitem[\protect\citeauthoryear{{Rees}}{{Rees}}{1982}]{rees1982}
{Rees} M.~J.,  1982, in {Riegler} G.~R.,  {Blandford} R.~D.,  eds,  American
  Institute of Physics Conference Series Vol. 83, The Galactic Center. pp
  166--176, \mn@doi{10.1063/1.33482}

\bibitem[\protect\citeauthoryear{{Saikia}}{{Saikia}}{2022}]{2022JApA...43...97S}
{Saikia} D.~J.,  2022, \mn@doi [Journal of Astrophysics and Astronomy]
  {10.1007/s12036-022-09863-2}, \href
  {https://ui.adsabs.harvard.edu/abs/2022JApA...43...97S} {43, 97}

\bibitem[\protect\citeauthoryear{{Shakura} \& {Sunyaev}}{{Shakura} \&
  {Sunyaev}}{1973}]{Shakura1973}
{Shakura} N.~I.,  {Sunyaev} R.~A.,  1973, \aap, \href
  {http://adsabs.harvard.edu/abs/1973A%26A....24..337S} {24, 337}

\bibitem[\protect\citeauthoryear{{Smol{\v c}i{\'c}} et~al.,}{{Smol{\v c}i{\'c}}
  et~al.}{2017}]{smolcic17}
{Smol{\v c}i{\'c}} V.,  et~al., 2017, \mn@doi [\aap]
  {10.1051/0004-6361/201730685}, \href
  {http://adsabs.harvard.edu/abs/2017A%26A...602A...6S} {602, A6}

\bibitem[\protect\citeauthoryear{{Spinoglio} \&
  {Fern{\'a}ndez-Ontiveros}}{{Spinoglio} \&
  {Fern{\'a}ndez-Ontiveros}}{2021}]{2021IAUS..356...29S}
{Spinoglio} L.,  {Fern{\'a}ndez-Ontiveros} J.~A.,  2021, in {Povi{\'c}} M.,
  {Marziani} P.,  {Masegosa} J.,  {Netzer} H.,  {Negu} S.~H.,   {Tessema}
  S.~B.,  eds,  IAU Symposium Vol. 356, Nuclear Activity in Galaxies Across
  Cosmic Time. pp 29--43 (\mn@eprint {arXiv} {1911.12176}),
  \mn@doi{10.1017/S1743921320002549}

\bibitem[\protect\citeauthoryear{{Timmerman} et~al.,}{{Timmerman}
  et~al.}{2022}]{2022A&A...658A...5T}
{Timmerman} R.,  et~al., 2022, \mn@doi [\aap] {10.1051/0004-6361/202140880},
  \href {https://ui.adsabs.harvard.edu/abs/2022A&A...658A...5T} {658, A5}

\bibitem[\protect\citeauthoryear{{Tremaine} et~al.,}{{Tremaine}
  et~al.}{2002}]{2002ApJ...574..740T}
{Tremaine} S.,  et~al., 2002, \mn@doi [\apj] {10.1086/341002}, \href
  {https://ui.adsabs.harvard.edu/abs/2002ApJ...574..740T} {574, 740}

\bibitem[\protect\citeauthoryear{{Villarroel} \& {Korn}}{{Villarroel} \&
  {Korn}}{2014}]{2014NatPh..10..417V}
{Villarroel} B.,  {Korn} A.~J.,  2014, \mn@doi [Nature Physics]
  {10.1038/nphys2951}, \href
  {https://ui.adsabs.harvard.edu/abs/2014NatPh..10..417V} {10, 417}

\bibitem[\protect\citeauthoryear{{Whittam} et~al.,}{{Whittam}
  et~al.}{2022}]{2022MNRAS.516..245W}
{Whittam} I.~H.,  et~al., 2022, \mn@doi [\mnras] {10.1093/mnras/stac2140},
  \href {https://ui.adsabs.harvard.edu/abs/2022MNRAS.516..245W} {516, 245}

\bibitem[\protect\citeauthoryear{{Willott}, {Rawlings}, {Blundell}  \&
  {Lacy}}{{Willott} et~al.}{1999}]{1999MNRAS.309.1017W}
{Willott} C.~J.,  {Rawlings} S.,  {Blundell} K.~M.,   {Lacy} M.,  1999, \mn@doi
  [\mnras] {10.1046/j.1365-8711.1999.02907.x}, \href
  {https://ui.adsabs.harvard.edu/abs/1999MNRAS.309.1017W} {309, 1017}

\bibitem[\protect\citeauthoryear{{Zaja{\v{c}}ek} et~al.,}{{Zaja{\v{c}}ek}
  et~al.}{2019}]{2019A&A...630A..83Z}
{Zaja{\v{c}}ek} M.,  et~al., 2019, \mn@doi [\aap]
  {10.1051/0004-6361/201833388}, \href
  {https://ui.adsabs.harvard.edu/abs/2019A&A...630A..83Z} {630, A83}

\bibitem[\protect\citeauthoryear{{Zhang}, {Fan}  \& {Zhu}}{{Zhang}
  et~al.}{2021}]{2021PASJ...73..313Z}
{Zhang} L.,  {Fan} J.,   {Zhu} J.,  2021, \mn@doi [\pasj]
  {10.1093/pasj/psaa122}, \href
  {https://ui.adsabs.harvard.edu/abs/2021PASJ...73..313Z} {73, 313}

\makeatother
\end{thebibliography}



\appendix


\bsp	
\label{lastpage}
\end{document}